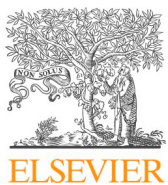
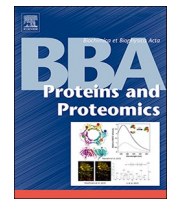

# The impact of physiological stress conditions on protein structure and trypsin inhibition of serine protease inhibitor Kazal type 1 (SPINK1) and its N34S variant

Ina Buchholz[a,b], Felix Nagel[a,b], Annelie Klein[a,b], Preshit R. Wagh[c], Ujjwal M. Mahajan[c,d], Andreas Greinacher[e], Markus M. Lerch[c], Julia Mayerle[c,d], Mihaela Delcea[a,b,⁎]

[a] *Institute of Biochemistry, University of Greifswald, 17489 Greifswald, Germany*
[b] *ZIK HIKE, Center for Innovation Competence "Humoral Immune Reactions in Cardiovascular Diseases", University of Greifswald, 17489 Greifswald, Germany*
[c] *Department of Medicine A, University Medicine Greifswald, 17475 Greifswald, Germany*
[d] *Department of Medicine II, Ludwig-Maximilian University of Munich, 81377 Munich, Germany*
[e] *Institute for Immunology and Transfusion Medicine, University Medicine of Greifswald, 17475 Greifswald, Germany*

ABSTRACT

One of the most common mutations in the serine protease inhibitor Kazal type 1 (*SPINK1*) gene is the N34S variant which is strongly associated with chronic pancreatitis. Although it is assumed that N34S mutation constitutes a high-risk factor, the underlying pathologic mechanism is still unknown. In the present study, we investigated the impact of physiological stress factors on SPINK1 protein structure and trypsin inhibitor function using biophysical methods. Our circular dichroism spectroscopy data revealed differences in the secondary structure of SPINK1 and N34S mutant suggesting protein structural changes induced by the mutation as an impairment that could be disease-relevant. We further confirmed that both SPINK1 ($K_D$ of 0.15 ± 0.06 nM) and its N34S variant ($K_D$ of 0.08 ± 0.02 nM) have similar binding affinity and inhibitory effect towards trypsin as shown by surface plasmon resonance and trypsin inhibition assay studies, respectively. We found that stress conditions such as altered ion concentrations (i.e. potassium, calcium), temperature shifts, as well as environmental pH lead to insignificant differences in trypsin inhibition between SPINK1 and N34S mutant. However, we have shown that the environmental pH induces structural changes in both SPINK1 constructs in a different manner. Our findings suggest protein structural changes in the N34S variant as an impairment of SPINK1 and environmental pH shift as a trigger that could play a role in disease progression of pancreatitis.

## 1. Introduction

Serine protease inhibitor Kazal type 1 (SPINK1) also known as pancreatic secretory trypsin inhibitor (PSTI) binds to the proteolytic enzyme trypsin in the pancreas and inhibits its activity preventing autodigestion of the surrounding tissues by uncontrolled, premature activation of trypsinogen and other zymogens. SPINK1 is produced in acinar cells of the pancreas as a 79 amino acid protein including a 23 residue signal peptide sequence [1,2], which is cleaved before storage in zymogen granules. Further, this 6.2 kDa inhibitor is secreted to the pancreatic juice along with the digestive zymogens [3]. Cationic trypsin is the most abundant isoform of trypsin in pancreatic juice [4]. SPINK1 interacts specifically with cationic trypsin as a 1:1 complex [5] mediated through its reactive site residue K41 via competitive inhibition mimicking the protease substrate. This is also known as "Laskowski mechanism", which is shared with many other protease inhibitors [6]. SPINK1 bound to trypsin is cleaved at its reactive site, though with low catalytic efficiency making proteolysis of the inhibitor very slow in comparison to actual trypsin substrates. As shown in Fig. 1A, SPINK1 globular peptide structure was investigated by X-ray crystallography [7,8] and reveals a structure shared by all classical Kazal inhibitors. A short central alpha-helix, as well as an antiparallel beta-sheet are surrounded by random coil and loop regions. The protein lacks glycosylation sites and its structure is stabilized by three intramolecular disulfide bonds (yellow) at positions C32/C61, C39/C58 and C47/C79.

Several mutations of SPINK1 mature peptide N34S, G48E, D50E, Y54H, P55S, R65Q and R67C are known [9,10]. The N34S mutation, whose location is depicted in red in Fig. 1A and B, is strongly associated with chronic pancreatitis [9] and represents the most common mutant of SPINK1 appearing in 13–25% of chronic pancreatitis patients, but also in up to 1.5% of healthy population [9–17]. The percentage of healthy population carrying this particular mutation is quite large considering a prevalence of chronic pancreatitis of 0.02%. Frequently, pancreatitis processes idiopathically [18]. Potential etiologies include toxins, infections [19], medications [20], autoimmune disorders [21], vascular causes [22], or anatomic and functional causes [18].

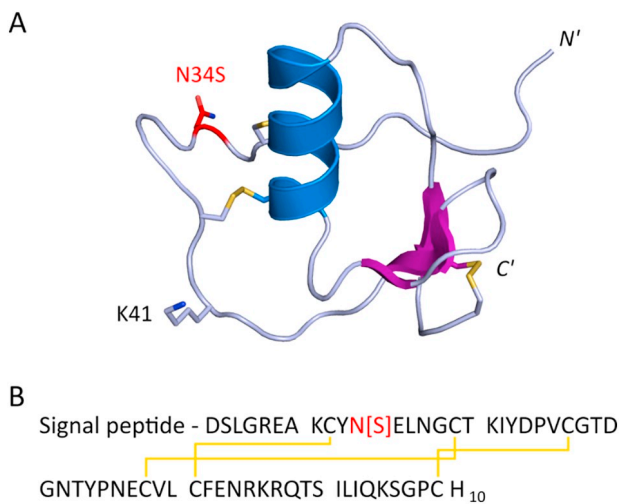

**Fig. 1.** (A) X-ray structure of SPINK1 shown as cartoon model with N34S mutational site (red) and disulfide bonds (yellow) [7], PDB-ID 1hpt. The reactive site residue K41 is depicted in grey. PyMOL 2.0 software was used to create this section. (B) Primary amino acid sequence of SPINK1 and N34S mutant as used in this study. The N34S mutational site is colored in red, whereas disulfide bonds are indicated in yellow.

According to current knowledge, SPINK1 and N34S mutant have similar binding affinity and inhibitory effect towards trypsin as well as unaffected expression and secretion levels [23–28]. Pfützer et al. assumed that the mutation alone does not cause pancreatitis, but potentially acts as a disease modifier [10], creating an increased risk for chronic pancreatitis [29–31]. The underlying pathologic mechanism is still unknown. A trigger (e.g. a bacterial or viral infection, a previous inflammatory disorder provoking an increased body temperature, ion concentration fluctuations or shifts in environmental pH) might be needed in addition to the N34S mutation to induce pathogenicity. In this regard, we have recently shown that another endogenous, secretable protein, the chemokine platelet factor 4 (PF4) with a MW of 7.8 kDa undergoes conformational changes depending on its environment [32–34]. This conformational change triggers clustering, especially in the presence of polyanions, and an autoimmune response [35].

In the present study, we characterized SPINK1 and its N34S variant as well as the interaction with human cationic trypsin under various stress conditions that could induce changes in SPINK1 protein structure to assess whether these changes might be a potential mechanism for the increased prevalence of pancreatitis in individuals carrying this mutation.

## 2. Material & methods

Unless otherwise stated, all chemicals were purchased from Sigma (Taufkirchen, Germany).

### 2.1. Cell culture and transfection

Human embryonic kidney 293 T cells (HEK293T) were cultured in T75 cell culture flasks with Dulbecco's Modified Eagles Medium (DMEM; Biowest, Nuaillé, France) with additional supplements 10% fetal calf serum, 4 mM L-glutamine and 1% penicillin/streptomycin at 37 °C in a humidified atmosphere containing 5% $CO_2$. Transfection was carried out 18–24 h after cell division. For transfection, a mixture of 30 μg plasmid DNA (pcDNA 3.1 (−) SPINK1 wild type and N34S mutant sequence including a C-terminal $His_{10}$-tag) and 80 μL transfection reagent TransIT-LT1 (MoBiTec, Göttingen, Germany) was prepared in 2 mL Reduced-Serum Medium OptiMEM (Life Technologies, Carlsbad, US). The mixture was incubated for 30 min at room temperature (RT), before addition to the DMEM cell culture medium. The transfected cell culture was incubated for 4 h at 37 °C and the DMEM medium was replaced by OptiMEM medium. Protein expression was carried out for 72 h starting from the point of medium change.

### 2.2. SPINK1 purification

The cell culture medium containing the secreted SPINK1 or N34S mutant (MW: 7.6 kDa) was harvested and centrifuged at RT for 30 min at 4750g. For purification of both constructs by $Ni^{2+}$ affinity chromatography, the supernatant was loaded on a HisTrap excel column (GE Healthcare, Uppsala, Sweden) with a flow rate of 1 mL/min and the column was washed with phosphate buffer (20 mM $NaH_2PO_4$, 500 mM NaCl, pH 7.4) containing 20 mM imidazole. An additional washing step using 60 mM imidazole was included to remove unspecifically bound proteins. The elution was carried out by phosphate buffer with 500 mM imidazole. The eluted protein was concentrated using centrifugal filter units with a cut-off of 3 kDa (Merck, Darmstadt, Germany) and dialyzed in 3.5 kDa cut-off cassettes (Slide-A-Lyzer Mini Dialysis Device; Thermo Fisher, Darmstadt, Germany) overnight at 4 °C containing the respective buffer. Protein purity was verified by silver stained non-reductive SDS-PAGE and protein concentration was determined by bicinchoninic acid assay kit (Sigma).

### 2.3. Circular dichroism (CD) spectroscopy

CD spectra were acquired with a Chirascan CD spectrometer (Applied Photophysics, Leatherhead, UK) equipped with a temperature control unit (Quantum Northwest, Liberty Lake, US) at 25 °C. Measurements in the far-UV region (195–260 nm) were performed with a 5 mm path length cuvette (110-QS; Hellma Analytics, Müllheim, Germany) and a protein concentration of 6.5 μM (50 μg/mL). Spectra were recorded with a bandwidth of 1.0 nm, a scanning speed of 15 nm/min and five repetitions. During data analysis, all spectra were blank- and volume-corrected. The estimation of the secondary structural content was carried out via deconvolution of CD spectra by the software CDNN using a database of 33 reference proteins [36]. For titration with different ions, SPINK1 was prepared in 10 mM Tris, 1 mM NaCl, pH 8.0 after purification. After addition of either 1 mM or 5 mM $CaCl_2$ or KCl solution, SPINK1 or pure buffer (blank titration) were equilibrated for 30 min before CD measurement. Temperature-dependent CD spectra of SPINK1 were recorded at 25 °C, 85 °C and after thermal denaturation (130 °C for 15 min) at 25 °C in 10 mM Tris, 1 mM $CaCl_2$, pH 8.0. pH titrations (pH 8.0–3.0) were performed starting from either the protein in 10 mM Tris, 1 mM $CaCl_2$, pH 8.0 or pure buffer and aliquots of 0.5 or 1% HCl solution were added stepwise. The pH was measured directly in the cuvette after an equilibration time of 10 min followed by recording of the CD spectra. For CD melting experiments 6.5–10.5 μM (50–80 μg/mL) SPINK1 in 10 mM Tris, 1 mM $CaCl_2$, pH 8.0 was scanned from 20 to 85 °C with a heating rate of 18 °C/h. CD signal was recorded with a step size of 0.5 °C (± 0.1 °C tolerance) at single wavelength of 206 and 218 nm. The temperature sensor was placed directly in the cuvette containing protein or buffer, respectively.

### 2.4. Surface plasmon resonance (SPR) spectroscopy

Kinetic analyses were performed using a Biacore T200 (GE Healthcare) at 25 °C. Amine coupling chemistry (NHS/EDC kit, GE Healthcare) was used to covalently immobilize SPINK1 and N34S mutant on a CM5 sensor chip surface (GE Healthcare). SPINK1 constructs were immobilized using a concentration of 10 μg/mL in 10 mM MES (2-

(*N*-morpholino)ethanesulfonic acid) buffer, pH 5.5 to a density below 40 response units (RU). Ethanolamine-inactivated flow cells were used for referencing and data was collected at 10 Hz. Human cationic trypsin (MW: 26.5 kDa; GenWay Biotech, San Diego, US) was prepared as two-fold dilutions in running buffer (100 mM Tris, 150 mM NaCl, 1 mM $CaCl_2$, 0.05% Tween20, pH 8.0 or 4.8, respectively). Trypsin was injected over the biosensor surface for 180 s followed by a dissociation time of 1000 s at a flow rate of 40 μL/min. The surfaces were regenerated after each cycle with regeneration solution (10 mM glycine/HCl, pH 2.0) for 100 s. Three independent experiments were carried out in each case. Data from six different concentrations between 1.56 and 50 nM were double referenced and fitted globally by a heterogeneous ligand model (BIAevaluation 3.1, GE Healthcare) to determine the binding parameters $K_{D1,2}$, $k_{on1,2}$ and $k_{off1,2}$.

### 2.5. Trypsin inhibition assay (TIA)

11 different SPINK1 concentrations (0–100, 400 nM) were prepared via serial dilution in assay buffer (100 mM Tris, 1 mM $CaCl_2$, pH 8.0 with 0.05% Tween20 to avoid protein surface adsorption, initial conditions). In order to test stress conditions during TIA, single parameters of the assay buffer were varied, namely 1 mM or 5 mM KCl instead of $CaCl_2$, or 5 mM $CaCl_2$, or the buffer was adjusted to pH 6.5, 4.8 or 3.0. Further, incubation and assay temperature were set to 45 °C or thermally denatured (130 °C for 15 min) SPINK1 was used in TIA. Samples were placed in triplicate (10 μL/well) on a black microtiter plate (Cat. #655209; Greiner bio-one, Frickenhausen, Germany). 10 μL of 100 nM human cationic trypsin was added to each well and incubated for 30 min at RT under mild shaking. 80 μL of 25 μM trypsin rhodamine 110 substrate (Thermo Fisher) in the respective buffer was added to each well and the fluorescence ($\lambda_{ex}$ 485, $\lambda_{em}$ 521) was detected at 37 °C with data point record each 30 s. For TIA data analysis, the slope (linear range typically 0–1000 s, but for pH 4.8 condition 0–4800 s) was calculated. Final plots consist of ≥9 independent reproductions and show % trypsin inhibition vs. SPINK1 concentration with standard deviation error bars, whereas trypsin activity without SPINK1 at the respective stress condition was used as 0% and with 400 nM SPINK1 as 100% reference for inhibition.

### 2.6. Differential scanning calorimetry (DSC)

DSC measurements were carried out with a PEAQ-DSC (Malvern Instruments, Herrenberg, Germany). 60 μM (450 μg/mL) SPINK1 in Hepes buffer (10 mM Hepes, 50 mM NaCl, pH 8.0) was scanned from 25 to 130 °C with a heating rate of 60 °C/h. None or low feedback mode was used. Determination of the protein transition temperature was done after buffer baseline subtraction. DSC rescans with the same parameters were recorded to prove irreversible denaturation of SPINK1.

## 3. Results

### 3.1. SPINK1 and N34S exhibit differences in protein secondary structure

CD spectroscopy studies of both SPINK1 constructs (Fig. 2) showed that SPINK1 (black) exhibits the characteristic CD profile as published earlier [37]. Compared to SPINK1, the N34S mutant (red) shows differences in protein secondary structure, remarkable around the shoulder of the spectrum (220 nm) and at the minimum at 206 nm. CD data deconvolution revealed a slight increase in alpha-helical content and decrease in antiparallel beta-sheet after N34S mutation (Fig. S1).

### 3.2. Consistent trypsin binding affinity of SPINK1 and N34S

SPR experiments were carried out in order to determine the dissociation constant ($K_D$) as well as the kinetic binding parameters of SPINK1 and N34S mutant interacting with human cationic trypsin

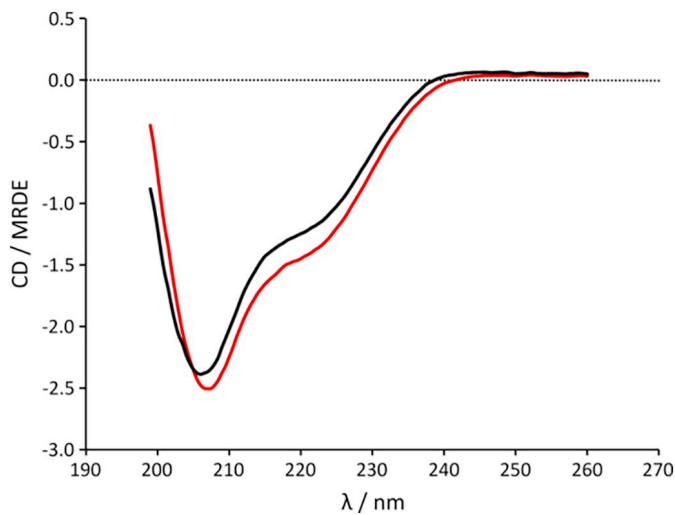

**Fig. 2.** Representative CD spectra of SPINK1 (black) and N34S mutant (red) showing differences in secondary protein structure. Spectra were recorded in 10 mM Tris, 1 mM $CaCl_2$, pH 8.0 at 25 °C.

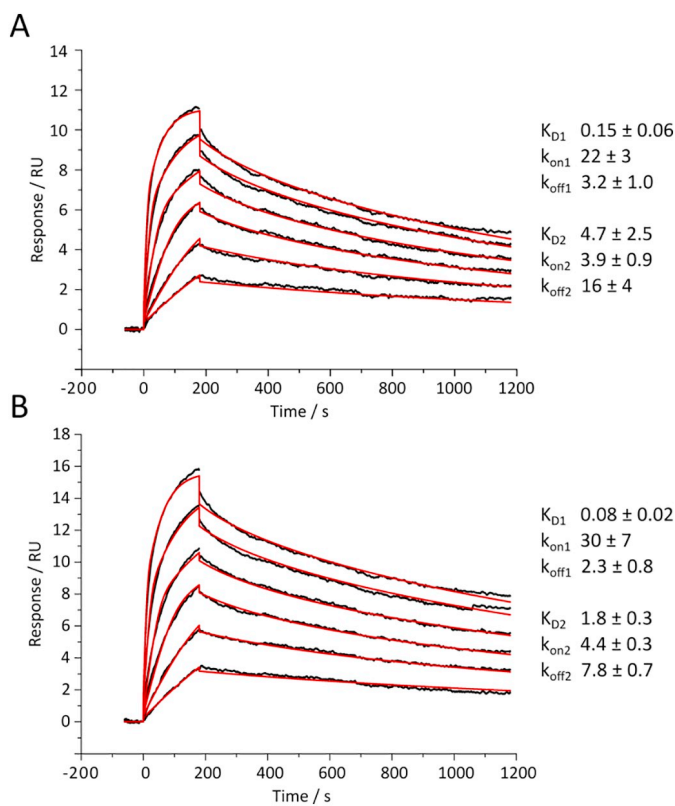

**Fig. 3.** Representative SPR sensorgrams of human cationic trypsin interacting with SPINK1 (A) or N34S mutant (B) immobilized on the sensor chip surface. Experiments were carried out in 100 mM Tris, 150 mM NaCl, 1 mM $CaCl_2$, 0.05% Tween20, pH 8.0 at 25 °C. Trypsin was injected at concentrations of 50, 25, 12.5, 6.25, 3.13 and 1.56 nM. The reference subtracted sensorgrams are shown in black and the heterogeneous ligand model fit in red. The dissociation constants $K_{D1,2}$ [$10^{-9}$ M] as well as the kinetic binding parameters $k_{on1,2}$ [$10^5$ 1/Ms], $k_{off1,2}$ [$10^{-4}$ 1/s] with standard deviation of three independent measurements are summarized next to the respective sensorgram (see also Table S1).

(Fig. 3). SPINK1 (Fig. 3A) or N34S mutant (Fig. 3B) was immobilized on the SPR sensor chip surface and trypsin was used as analyte. Binding parameters determined from SPR data fit analyses with a heterogeneous ligand model are summarized next to the sensorgrams in Fig. 3. Both SPINK1 constructs revealed $K_D$ in the range of 0.1 nM for a high affinity binding site ($K_{D1}$) and 3 nM for a low affinity binding site ($K_{D2}$) at pH 8.0. Accordingly, binding rate constants are comparable among both constructs. Thus, no significant differences in the binding affinity of SPINK1 and N34S mutant towards trypsin could be found. However, the high affinity sites display a significant larger association rate constant $k_{on}$ (i.e. more rapid complex formation) and a smaller dissociation rate constant $k_{off}$, which indicates increased complex stability (i.e. slower complex dissociation) compared to the low affinity sites.

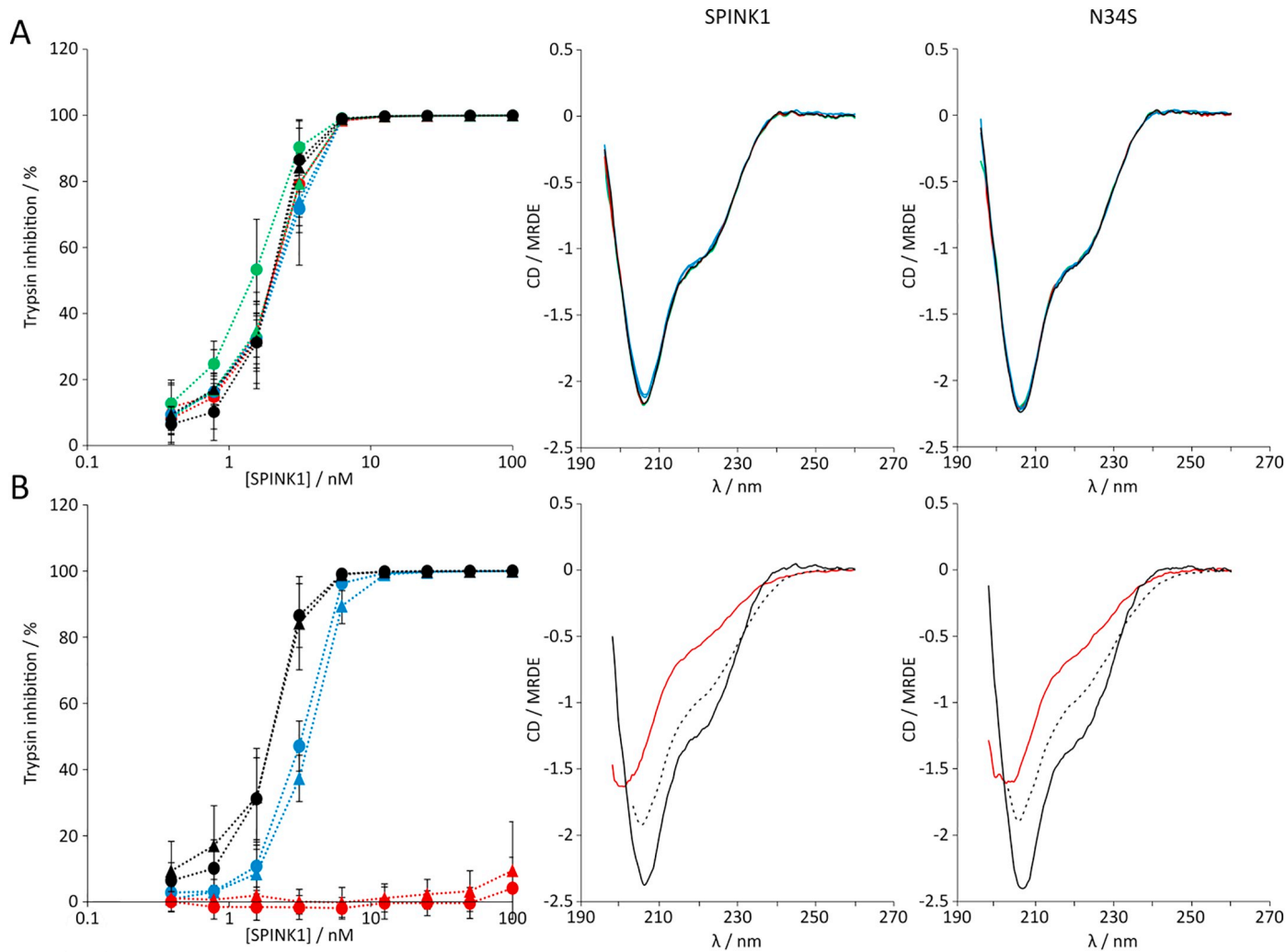

**Fig. 4.** TIA results and CD spectra of SPINK1 and N34S mutant under stress conditions. SPINK1 (spheres) and N34S mutant (triangles) with standard assay conditions containing 1 mM CaCl$_2$, pH 8.0 (black). In the left panels the TIA is shown, whereas middle and right panels display representative CD spectra under stress conditions of the SPINK1 and N34S mutant, respectively. (A) SPINK1 and N34S mutant at different K$^+$ and Ca$^{2+}$ ion concentrations, including 1 mM (blue) and 5 mM (green) K$^+$, as well as 1 mM (standard condition, black) and 5 mM (red) Ca$^{2+}$. (B) SPINK1 and N34S mutant at different temperatures. In addition to the standard conditions at 37 °C, the TIA results at 45 °C (blue) and after thermal denaturation (red) are shown. CD spectra of SPINK1 and N34S mutant at 25 °C (black), 85 °C (black, dotted) and after thermal denaturation (red) are displayed.

Further, SPINK1 and N34S mutant were investigated under certain stress conditions including increasing potassium and calcium ion concentrations, temperature shifts, and at physiological occurring pH range.

### 3.3. Effect of ions and temperature on the structure and function of SPINK1 and N34S

The ability of SPINK1 and N34S mutant to inhibit human cationic trypsin under different conditions was monitored using a trypsin inhibition assay (TIA). The inhibitor effect of both proteins towards trypsin is very similar (left panels in Fig. 4A and B, black spheres for SPINK1 and black triangles for N34S) under standard assay conditions (1 mM Ca$^{2+}$, pH 8.0, 37 °C). At equimolar ratios of SPINK1 and trypsin, almost no trypsin activity is observed, which shows a very tight binding. Furthermore, these results support the structural integrity and activity of our SPINK1 constructs and confirm the uniform inhibitory profile of SPINK1 and N34S mutant under these conditions, as reported earlier [23,24].

Several groups investigated the presence and importance of ion channels and transporters in the membrane of zymogen granules controlling ion concentrations (e.g. Ca$^{2+}$, K$^+$, H$_3$O$^+$) [38]. In Fig. 4A (left panel), the TIA for SPINK1 and N34S mutant with different potassium and calcium ion concentrations are shown. Low potassium ion concentrations (1 mM, blue) as well as varying calcium ion concentrations (5 mM, red) do not alter the inhibitory effect of SPINK1 and N34S mutant in comparison to the initial conditions (black). However, a potassium ion concentration of 5 mM (green) appears to increase the inhibitory effect of SPINK1, but not of the N34S mutant.

In order to elucidate whether this change in trypsin inhibitory effect could result from a change in SPINK1 secondary protein structure, CD spectra with rising potassium or calcium ion concentrations were recorded (middle and right panels in Fig. 4A). It was found that SPINK1 and N34S mutant completely conserve their secondary structure under the increasing potassium and calcium ion concentrations tested, as indicated by the absence of changes in CD spectra.

To investigate the influence of increasing temperatures towards the trypsin inhibition, we studied SPINK1 and N34S mutant in TIA at 45 °C (left panel, blue color in Fig. 4B). The inhibitory effect of SPINK1 is decreased under increasing temperature, requiring higher concentrations of SPINK1 to maintain inhibition. After thermal denaturation of both SPINK1 constructs, TIA displays a complete loss of inhibitory effect (red). Furthermore, SPINK1 and N34S mutant possess identical inhibitory profiles for the tested temperature states. CD spectra (middle and right panels in Fig. 4B) at 25 °C (black), 85 °C (black, dotted) and after thermal denaturation (red) show that SPINK1 secondary structures are modified at 85 °C. However, after thermal denaturation, a distinct loss of structure is visible, which was not accomplished at 85 °C. In the latter case, the SPINK1 structural change is completely reversible (data not shown), which demonstrates high thermal stability for both SPINK1 and N34S mutant.

Furthermore, the thermal stability of the SPINK1 structures was explored by single-wavelength CD melting experiments (20–85 °C, Fig. S2). Within the physiological body temperature range SPINK1 does not offer significant structural changes including SPINK1 structure at 45 °C. CD melting profiles do not exhibit a transition point, but they rather show a continuous and linear decrease in CD signal recorded at 206 (minimum) and 218 nm (shoulder), which constitute the crucial points to observe for SPINK1 structural changes. In addition, DSC profiles (25–130 °C) were recorded to observe SPINK1 thermal stability over a wide temperature range (Fig. S3). In agreement with CD melting data, a very broad profile is observed, showing a transition point around 90–95 °C for both constructs. This profile indicates protein transition with very low cooperativity, which is typical for low-structured proteins and was already reported for other SPINK1 constructs with concordant melting points [37]. After DSC melting, SPINK1 structure is irreversibly impaired.

### 3.4. Effect of environmental pH stress on the structure and function of SPINK1 and N34S

As a last stress condition, SPINK1 and N34S mutant were exposed to different environmental pH within a physiological range. In addition to the initial pH 8.0 (black), acidic conditions of pH 6.5 (blue) and 4.8 (red) were tested in TIA (left panel in Fig. 5). At pH 6.5, SPINK1 displays the same inhibitory effect compared to standard conditions (pH 8.0), whereas the N34S mutant shows a minor decrease in inhibition. At pH 4.8, the typical sigmoidal inhibition profile is changed towards a reduced slope and inhibitory effects of SPINK1 and N34S mutant are decreased. As for the pH 6.5 conditions, the N34S mutant shows slightly less inhibition compared to SPINK1 at pH 4.8. TIA results at pH 3.0 did not indicate any substrate conversion even in the absence of SPINK1 (data not shown).

In Fig. 5, CD spectra of pH titrations for SPINK1 (middle panel) and N34S mutant (right panel) are shown. Both proteins exhibit significant changes in their CD spectra between pH 8.0 and 3.0. While the CD signal of the shoulder (222 nm) decreases continuously, the minimum (207 nm) shows a reversible behavior. The amplitude of the CD signal at 207 nm is reduced until pH 6 and increases again at lower pH forming a triangle-shaped structural profile. Although similar secondary structural changes within the pH shift are found for SPINK1 and N34S mutant, the mutant presents a faster regain of CD signal (insert in the right panel of Fig. 5). SPINK1 secondary structure is sensitive to pH changes within the physiological range and a comparison to CD data of thermal denatured SPINK1 and TIA results demonstrates that pH-treatment does not induce protein denaturation.

TIA and CD spectroscopy experiments revealed that low potassium and varying calcium ion concentrations as well as physiological temperatures do not result in differences in trypsin inhibition of SPINK1 and N34S mutant or changes in protein structure. For an increased potassium ion concentration (5 mM) a divergent behavior of SPINK1 and N34S mutant was shown in TIA, which does not correlate to changes in protein structure. Nevertheless, the environmental pH triggers structural changes and trypsin inhibition for both SPINK1 constructs in a different manner, which may play a crucial role in pancreatitis.

### 3.5. Trypsin binding of SPINK1 and N34S at acidic pH

In order to verify our pH-dependent TIA results and to check whether trypsin inhibition could be reduced for the N34S mutant at acidic pH compared to SPINK1, a respective SPR binding study was carried out. The corresponding sensorgrams for trypsin binding at pH 4.8 are shown in Fig. 6. Association and dissociation phases are highly similar for both constructs over a broad concentration range of injected trypsin. Hence, similar binding affinities are qualitatively suggested. SPINK1 and N34S mutant revealed dissociation constants $K_D$ in the range of 1 nM for a high affinity binding site ($K_{D1}$) and 25 nM for a low affinity binding site ($K_{D2}$) at pH 4.8. $k_{on}$ is comparable for both constructs and affinity sites, whereas $k_{off}$ is one order of magnitude higher for the low affinity sites. Consequently, the binding of SPINK1 to human cationic trypsin is tenfold decreased at pH 4.8 compared to the binding at pH 8.0. However, no significant differences in the binding affinity of the SPINK1 constructs could be found at acidic condition.

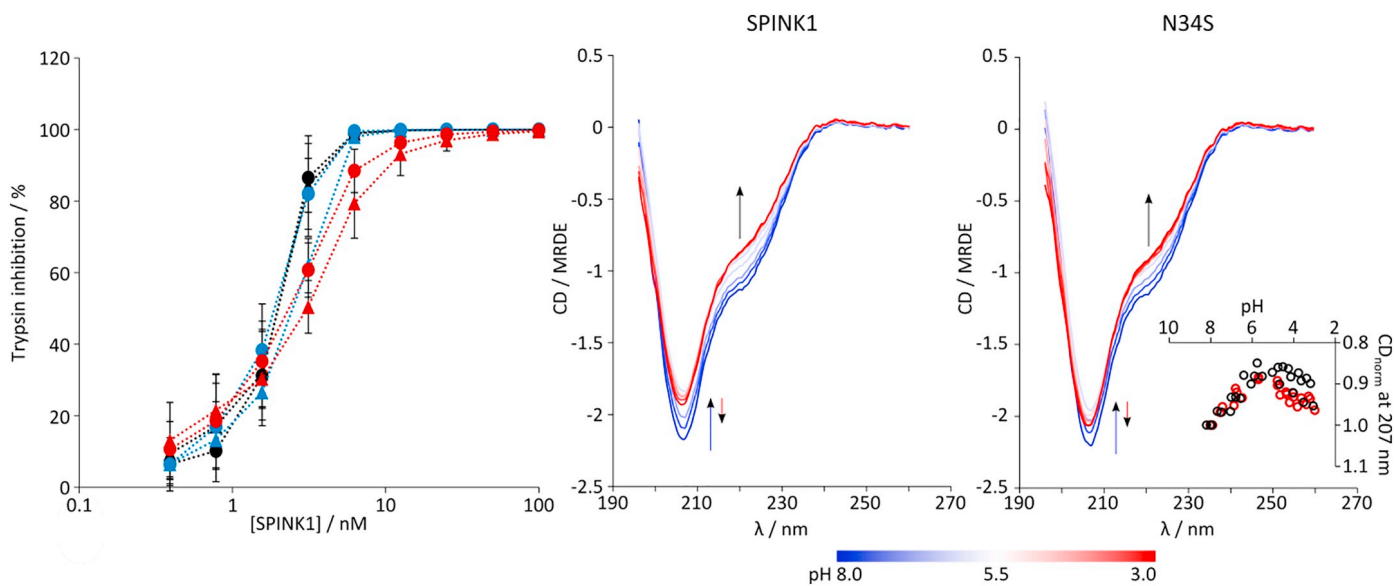

**Fig. 5.** TIA results and CD spectra of SPINK1 (spheres) and N34S mutant (triangles) under pH stress. In the left panel the TIA results at pH 8.0 (standard condition, black), pH 6.5 (blue) and pH 4.8 (red) are shown. Middle and right panels display representative CD spectra of pH titrations from pH 8.0 to pH 3.0 (colored blue to red) for SPINK1 and N34S mutant, respectively. The overall shift of the CD signal is indicated by arrows. The inset shows the normalized CD signal at 207 nm for SPINK1 (black) and N34S mutant (red) of four independent pH-titrations.

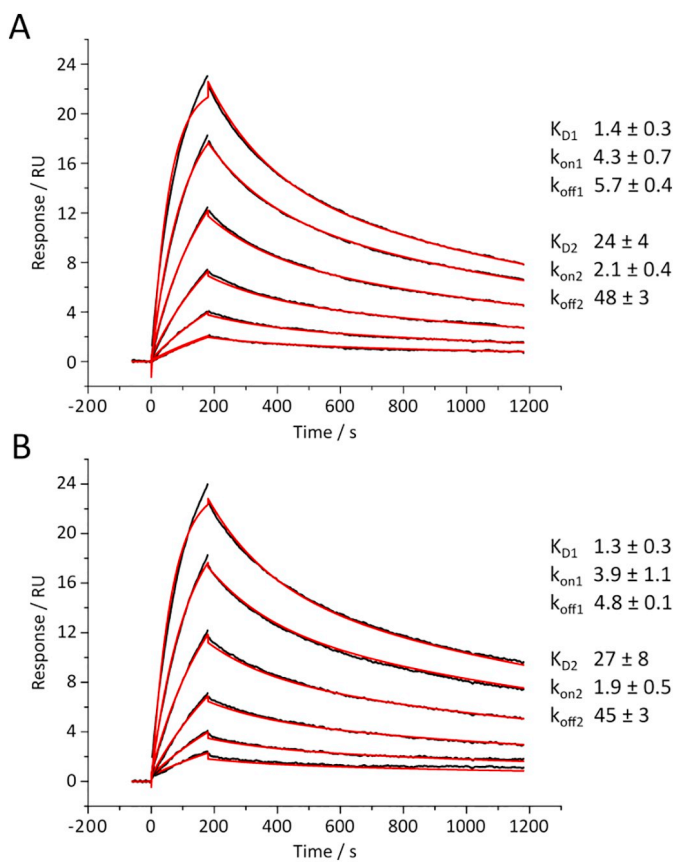

**Fig. 6.** Representative SPR sensorgrams of human cationic trypsin interacting with SPINK1 (A) or N34S mutant (B) immobilized on the sensor chip surface with similar immobilization levels. Experiments were carried out in 100 mM Tris, 150 mM NaCl, 1 mM $CaCl_2$, 0.05% Tween20, pH 4.8 at 25 °C. Trypsin was injected at concentrations of 50, 25, 12.5, 6.25, 3.13 and 1.56 nM. The reference subtracted sensorgrams are shown in black and the heterogeneous ligand model fit in red. The dissociation constants $K_{D1,2}$ [$10^{-9}$ M] as well as the kinetic binding parameters $k_{on1,2}$ [$10^5$ 1/Ms], $k_{off1,2}$ [$10^{-4}$ 1/s] with standard deviation are summarized next to the respective sensorgram (see also Table S1).

## 4. Discussion

This study reveals differences in the secondary structural composition between SPINK1 and N34S mutant. CD data deconvolution showed a slight increase in alpha-helical content and decrease in antiparallel beta-sheet after N34S mutation. Although, deconvolution constitutes a tool for rough secondary structure estimation, our predictions are in line with the findings of Kuwata et al. [39], who looked at primary amino acid sequence computer analysis of SPINK1 and N34S mutant. Their analysis suggested the elimination of a turn motif at N34/E35 while expanding the alpha-helical region. Furthermore, superposition with the structure of porcine SPINK1, which carries the S34 naturally, proposed hydrogen bond reorganization and local rearrangement of the peptide chain fold [10] [10]. Therefore, we consider that the N34S mutant has a minor folding defect, which might be a first impairment in comparison to the wild type protein and our CD spectroscopic analysis reveals the first experimental evidence for this structural differences between SPINK1 and its N34S variant.

Although the N34S mutant displays a minor folding defect, it leads to full trypsin inhibition as published previously [23,24]. These findings are in agreement with our SPR binding analysis of SPINK1 and N34S mutant to human cationic trypsin at pH 8.0 showing comparable dissociations constants $K_D$ for both constructs. Kiraly et al. deduced $K_D$ values from the influence on trypsin inhibition, which are in similar range (0.33 nM for SPINK1 and 0.26 nM for N34S) [24]. The heterogeneous ligand model has been previously used for the Kazal inhibitor SPINK9 [40] and represents a linear combination of two 1:1 L binding models describing two independent binding sites of the ligand. Since SPINK1 can exist in two states after complex formation with trypsin (one with an intact K41-I42 bond and one with a cleaved K41-I42) [41] it is reasonable to assume that the uncleaved and the cleaved* species display different kinetic properties. Consistent with other trypsin inhibitors [42], SPINK1 reveals a lower $K_{D1}$, whereas the second set of parameters refers to SPINK1* showing an increased $K_{D2}$, indicating a decreased binding affinity for the cleaved species. Our work provides the first direct and biophysical binding affinity study determining comparable $K_D$ of SPINK1 and N34S mutant under standard conditions.

Stress factors like potassium and calcium ion concentrations, as well as temperature shifts, were tested regarding their impact on SPINK1 inhibitory function and secondary structure. The inhibitory profiles of SPINK1 and N34S mutant are stable over a broad range of conditions. Within physiological temperatures, no changes in CD signal, but a decrease in inhibitory effect was observable, which did not differ between SPINK1 and N34S mutant. With the exception of potassium at 5 mM, ion concentration changes within the physiological range showed no differences between the inhibitory capacity of the SPINK1 WT and the N34S mutant. This finding is in agreement with previous studies investigating the influence of calcium on SPINK1 to trypsin interaction [23]. A slight change in inhibitory effect of SPINK1 was shown for high potassium concentrations (5 mM), but CD spectroscopy did not reveal any changes in secondary structure. Previously, it was assumed that different ions like calcium, potassium or zinc may play a role in pancreatitis [38]. In our study we did not find a correlation between altered SPINK1 inhibitor function and structural changes, although we cannot exclude that these ions interfere at other pathways leading to pancreatitis.

The physiological relevance of different structures of SPINK1 and N34S at varying environmental pH is high, as zymogen granules itself contain an acidic pH (approx. pH 5.8–6.5), whereas the cytosolic pH is 7.0–7.4 and the pancreatic juice has a pH of 8.0. A former study using TIA, reported that the residual trypsin activity is slightly higher in the case of the N34S mutant compared to SPINK1 at pH 7 and 9, but not at pH 5 and 8 [23]. Although minor differences in inhibitory effects between SPINK1 and N34S mutant at pH 6.5 and 4.8 were displayed in our TIA experiments, the standard deviation of single data points makes the significance of these differences questionable. Furthermore, SPR showed comparable binding affinities for both constructs at pH 4.8. We conclude that the differences in TIA may result from assay-based uncertainties, and SPINK1 and N34S mutant still present a reduced, but similar inhibitory profile at low pH. Moreover, it has been shown before that the rare SPINK1 variant R65Q shows full trypsin inhibitory effect, but remarkable loss of its secretion level, which is believed to result from degradation because of major misfolding [24,25,31], an effect also observed for PRSS1 mutations [43] [24,25,31]. Therefore, TIA is not suitable to monitor the impact of a possible trigger as a screening method because it is not sensitive enough for changes in SPINK1 structure and fails to account for other potential disease-causing mechanisms, as shown in the case of the R65Q mutation.

The tenfold increase in $K_D$ for the interaction of SPINK1 with trypsin at pH 4.8 compared to pH 8.0 cannot only be explained by the structural changes of SPINK1 visible in CD spectra, but also because the surface charge distribution of both binding partners is influenced by the pH. At acidic pH the net charge of the binding interface inverts and the negatively charged binding pocket for the positively charged K41 residue becomes positive itself which reduces the electrostatic interactions between trypsin and SPINK1.

Unlike the other stress conditions tested, shifts in environmental pH lead to changes in SPINK1 secondary structure. CD spectroscopy exhibited three structural states in the physiological pH range of pH 8.0–3.0 for both SPINK1 constructs with a maximal change of signal observed around pH 6. These states found in CD may correlate with altered properties of the protein at acidic pH with the N34S mutant being more prone to these changes compared to SPINK1. However, in comparison with the chemokine PF4, which undergoes major conformational changes, the conformational changes in SPINK1 and N34S mutant are minor.

## 5. Conclusion

Our biophysical study demonstrates differences in the secondary structure of SPINK1 and its N34S mutant, although they present similar binding affinity and inhibitory effect towards trypsin. It is important to note that the structural impairment of the N34S mutant, which was shown for the first time experimentally in this work, might be disease-relevant for pancreatitis. This is because the environmental conditions at which these changes occur (such as a pH of 4.8) are not only encountered within the relevant subcellular compartments of pancreatic acinar cells [44], but these compartments have also been shown to represent the initial site where premature protease activation begins [45]. We have shown that ions (i.e. potassium, calcium) and temperature shifts do not induce differences between SPINK1 and N34S mutant in trypsin inhibition and in protein structure, whereas environmental pH leads to structural changes, but not to altered trypsin inhibition among both constructs. However, minor changes in protein structure can have large impact on its physiological behavior. The environmental pH may act as a trigger which could alter the properties of this trypsin inhibitor under pathophysiological conditions by exposing new epitopes leading to pancreatitis. Thus, the N34S mutant might display pathogenicity towards a higher risk for pancreatitis through other mechanisms than by affecting trypsin inhibition. Moreover, we showed that, compared to trypsin inhibition assay which is not sensitive enough to monitor changes in SPINK1 structure, biophysical methods can be used to screen for possible disease triggers.

## Acknowledgments

We acknowledge Miklos Sahin-Toth for critical reading of the manuscript and Kristin J. Holl for technical support. This work was financially supported by the European Research Council (ERC) Starting Grant 'PredicTOOL' (637877) to M.D. and by the European Union (research project 'PePPP', grant number ESF/14-BM-A55-0047/16).

## Appendix A. Supplementary data

Supplementary data to this article can be found online at https://doi.org/10.1016/j.bbapap.2019.140281.